\documentclass[11pt]{article}
\usepackage[dvipsnames,x11names]{xcolor} 
\usepackage{lscape}
\usepackage{amssymb} 
\usepackage{graphicx}
\usepackage{footnote}
\usepackage[utf8]{inputenc}	
\usepackage{mathtools}
\usepackage{amsthm}
\usepackage{amsfonts}
\usepackage[english]{babel}
\usepackage{bigints}
\usepackage{mathrsfs}
\usepackage{lineno}
\usepackage{nicefrac}
\usepackage{enumerate}
\usepackage{bbm}

\usepackage{listings}

\usepackage{commath}
\usepackage{multirow}

\usepackage{newtxtext}
\usepackage[T1]{fontenc}

\usepackage{algorithm} 
\usepackage{algpseudocode} 
\usepackage{booktabs}

\usepackage[most]{tcolorbox}
\usepackage{mdframed}

\usepackage[]{natbib}
\usepackage{csquotes}

\usepackage{tabularx}

\usepackage[colorlinks=true,citecolor=blue,linkcolor=black,urlcolor=blue]{hyperref}
\definecolor{light-gray}{gray}{0.91}

\usepackage[margin=2cm]{geometry}

\usepackage{sectsty}
\sectionfont{\color{BrickRed}}

\sectionfont{\color{RoyalBlue4}}
\subsectionfont{\color{RoyalBlue4}}
\subsubsectionfont{\color{RoyalBlue4}}
\paragraphfont{\color{RoyalBlue4}}

\allowdisplaybreaks[1]
\usepackage{pifont}

\usepackage{lineno}

\makeindex

\def\dvg{\textrm{Div}}

\def\crl{\textrm{Curl }}

\newcommand{\imag}{\textbf{i}}

\newcommand{\Lagr}{\mathcal{L}}

\newcommand{\Crot}{\mathbb{C}_{\rm rot}}

\begin{document}

% \linenumbers

\begin{flushleft}

	{\huge\bfseries \color{DodgerBlue4}{A higher--order gradient framework for multi--scale seismic wave propagation}\par}

	\vspace{1.5em}  % Adds vertical space after the title
	
	% Authors on the same line:
	Rafael Abreu\textsuperscript{1,*}\\[0.5em]  % Author names with superscript affiliations
	
	% Affiliations:
	\scriptsize % Smaller font size for affiliation
	\textsuperscript{1} \textit{Institut de Physique du Globe de Paris, CNRS, Universit\'e de Paris, Paris, France} \\ [1em]  
	\textsuperscript{*} {email: rabreu@ipgp.fr}
	
\end{flushleft}

%	\footnotesize
%\noindent\textbf{Key Points}
%\begin{itemize}
%	\item Higher-order gradient elasticity provides a reduced dynamic homogenization framework for frequency-dependent seismic wave propagation in porous media.
%	
%	\item Common inertial and strain-gradient length scales govern P- and S-wave dispersion, while an additional curvature length scale controls shear-wave dispersion.
%	
%	\item The framework reproduces laboratory P- and S-wave dispersion in tight sandstones using only a small number of effective constitutive parameters.
%\end{itemize}

	\normalsize % Reset the font size for the date or other text

\begin{abstract}
	
	Modeling frequency-dependent seismic wave propagation in porous media remains a major challenge because classical poroelastic theories generally require a large number of poorly constrained constitutive parameters. We investigate seismic wave propagation within a higher-order gradient constitutive framework as an effective dynamic homogenization model for poroelastic media. The proposed formulation extends classical strain-gradient elasticity by incorporating an additional curvature contribution associated with rotational deformation, allowing independent descriptions of longitudinal and transverse wave dispersion within a unified continuum framework.
	
	A general dispersion relation is derived together with its macroscopic and microscopic velocity limits, providing a direct physical interpretation of the characteristic inertial and elastic length scales introduced by the model. The framework is validated against laboratory measurements of frequency-dependent P- and S-wave velocities in tight sandstones under different confining pressures. The inversion demonstrates that a common inertial characteristic length and a common strain-gradient characteristic length govern both wave types, while an additional curvature characteristic length is required only for shear waves.
	
	Comparison with classical Biot theory shows that the proposed framework reproduces the observed laboratory dispersion using only a small number of effective constitutive parameters, providing a reduced constitutive description of poroelastic wave propagation. These results suggest that higher-order gradient elasticity offers a physically interpretable and computationally efficient alternative for modeling frequency-dependent seismic wave propagation, with potential applications to seismic imaging, inversion, and natural hydrogen reservoir characterization.
	
\end{abstract}

%	\footnotesize
%\section*{Plain Language Summary}
%
%Seismic waves traveling through porous and fractured rocks are affected by pores, fluids, cracks, and other small-scale structures. Describing all of these effects explicitly generally requires many material properties that are difficult to determine from seismic observations. In this study, we investigate an alternative approach in which the influence of these small-scale structures is represented by a small number of effective characteristic lengths. The resulting model describes how seismic-wave velocities change with frequency and allows compressional and shear waves to respond differently to the internal structure of the material. We compare the model with laboratory measurements of compressional- and shear-wave velocities in tight sandstones and show that it reproduces the observed frequency-dependent behavior using only a few effective parameters. The results suggest that complex wave propagation in porous rocks can be represented without detailed knowledge of their pore geometry and fluid--solid interactions. This reduced description may be particularly useful for seismic imaging and inversion, where estimating a large number of poorly known material properties is difficult. \\

\newpage

\normalsize
\section{Introduction}

Seismic wave propagation in the Earth at local scales is naturally influenced by heterogeneities, porosity, fractures, and fluid-filled inclusions \citep{Aki2002,Mavko2009,sato2012seismic,montagner2026}. Despite important advances in theoretical seismology, incorporating poroelastic effects into practical seismic imaging remains a major challenge. On the one hand, laboratory observations of wave dispersion and attenuation can be reproduced by enriched poroelastic models that require a large number of constitutive parameters \citep[e.g.][]{kimura2006frame,kimura2008experimental,kimura2013shear}. On the other hand, estimating such a large parameter set from seismic observations alone is generally an ill-posed inverse problem, since many of these parameters are poorly constrained and cannot be independently measured from field experiments.

To overcome these limitations, we investigate seismic wave propagation within a higher-order gradient constitutive framework. Rather than explicitly modeling pore geometry or fluid--solid interactions, higher-order continuum theories enrich classical elasticity by introducing characteristic length scales that account for nonlocal interactions, wave dispersion, and size effects beyond classical linear elasticity \citep[e.g.][]{mindlin1965second,dell2017higher,pideri1997second,dell1998micro,cecchi2001heterogeneous,Askes2002,Askes2007}. The proposed framework extends the classical strain-gradient formulation by incorporating an additional curvature contribution associated with local rotations. This additional constitutive mechanism allows longitudinal and transverse waves to be modeled independently, providing sufficient flexibility to reproduce the different dispersive behaviors observed experimentally for both P and S waves.

Although higher-order continuum theories have been extensively studied over the last decades \citep{dell2017higher}, the physical interpretation of their constitutive parameters remains relatively limited for Earth sciences applications. We show that higher-order gradient elasticity can be interpreted as a dynamic homogenization framework for poroelastic wave propagation. Within this framework, the characteristic length scales acquire a direct mechanical interpretation through their relationship with the macroscopic and microscopic limits of seismic wave velocities and, at the same time, providing a reduced constitutive description of the effective influence of the microstructure on wave propagation. The proposed model reproduces the dominant poroelastic dispersion observed in laboratory experiments using only a reduced number of effective constitutive parameters, making it particularly attractive for large-scale seismic imaging and inversion.

This paper is organized as follows. In Section \ref{sec.Equations_of_Motion}, we present the higher-order gradient theoretical framework within a clear notation and with a particular emphasis to the physical interpretation of the model parameters. In Section \ref{se_GeneralExperimentalValidations}, we validate the model against experimental observations including frequency-dependent seismic velocities in tight sandstones. Section~\ref{sec.Discussion} provides a broader interpretation of the results. We analyze the implications of the model for frequency-dependent seismic responses. We further discuss possible connections with reservoir imaging. Finally, conclusions are drawn in Section~\ref{sec.Conclusions}.

\section{Theoretical foundations}
\label{sec.Equations_of_Motion}

We apply the Lagrangian formalism (also called the variational principle of least action and/or the principle of stationary action) to find equations of motion by minimizing the functional $S$ (principle of stationary action $\delta S = 0)$, defined as follows
\begin{align}
	S = \int_0^t \int _{\Omega} \Lagr \dif x \dif t = \int_0^t \int _{\Omega} (K - \mathcal{E}) \dif x \dif t ,
	\label{eq.Minimization_Functional}
\end{align}
were $\Omega$ denotes the volume of the continuum,  $\Lagr$ is the Lagrangian defined as the difference between kinetic energy $K$ and potential energy $\mathcal{E}$.

\subsection{Potential energy}

The introduced model belongs to the family of  gradient-elastic models \citep[e.g.][]{metrikine2002one,Askes2002,Askes2007}. The strain tensors $\varepsilon,\eta,	\kappa$ are given by the following expressions
\begin{align}
\begin{aligned}
	\varepsilon_{ij}
&=
\frac{1}{2}
\left(
\partial_j u_i
+
\partial_i u_j
\right),
\\
\eta_{ijk}
&=
\partial_k \varepsilon_{ij}
=
\frac{1}{2}
\left(
\partial_k\partial_j u_i
+
\partial_k\partial_i u_j
\right),
\\
\vartheta_i
&=
\frac{1}{2}
\epsilon_{ijk}\partial_j u_k,
\\
\kappa_{ij}
&=
\partial_j\vartheta_i
=
\frac{1}{2}
\epsilon_{ik\ell}
\partial_j\partial_k u_\ell ,
\end{aligned}
	\label{eq.Gradient_Strain_Measures}
\end{align}
where $u:\mathbb{R}^3 \rightarrow \mathbb{R}^3$ is the displacement vector.  
Assuming no coupling between the classical strain and the strain gradient, the elastic-energy density (potential energy) is given by
\begin{align}
	\mathcal{E}
	&=
	\frac{1}{2}
	\mathbb{C}_{ij\ell m}
	\varepsilon_{ij}\varepsilon_{\ell m}
	+
	\frac{1}{2}
	(L_c^2)_{kn}
	\mathbb{C}_{ij\ell m}
	\eta_{ijk}\eta_{\ell mn} +
	\frac{1}{2}
	L_d^2 
	(\Crot)_{ij\ell m}
	\kappa_{ij}\kappa_{\ell m }.
	\label{eq.Gradient_Elastic_Energy_General}
\end{align}
where $\mathbb{C}$ is the conventional symmetric fourth-order tensor of elastic constants \citep{Slawinski2015}, $\Crot$ is a fourth-order curvature elasticity tensor and $L_c^2$ is a symmetric second-order constitutive tensor with dimensions of length squared. 

The stress, double-stress and couple-stress tensors are given by the following expressions
\begin{align}
	\sigma_{ij}
	=
	\frac{\partial\mathcal{E}}
	{\partial\varepsilon_{ij}}
	=
	\mathbb{C}_{ij\ell m}\varepsilon_{\ell m},
	\qquad
	\tau_{ijk}
	=
	\frac{\partial\mathcal{E}}
	{\partial\eta_{ijk}}
	=
	(L_c^2)_{kn}
	\mathbb{C}_{ij\ell m}
	\eta_{\ell mn}, \quad m_{ij}= \frac{\partial\mathcal{E}}
	{\partial\kappa_{ij}}
	=
	L_d^2
	(\Crot)_{ij\ell m}
	\kappa_{\ell m}.
	\label{eq.Gradient_Stress_Tensors}
\end{align}

In isotropic media we can write 
\begin{align}
	\sigma_{ij}
	=
	\lambda\delta_{ij}\partial_\ell u_\ell
	+
	\mu
	\left(
	\partial_i u_j
	+
	\partial_j u_i
	\right),
	\quad
	\tau_{ijk}
	=
	 L_c^2 \, \partial_k \sigma_{ij},
	\quad 
	m_{ij}
	=
	\mu L_d^2\,\kappa_{ij},
	\label{eq.Isotropic_Gradient_Stresses}
\end{align}
where $\lambda,\mu$ are Lam\'e parameters.

\subsection{Kinetic energy and equations of motion}

Having defined the elastic free energy eq. \eqref{eq.Gradient_Elastic_Energy_General}, we define the corresponding kinetic energy as follows
\begin{align}
	K
	&=
	\frac{\rho}{2}
	\partial_t u_i\,\partial_t u_i
	+
	\frac{\rho L_t^2}{2}
	\partial_j\partial_t u_i\,
	\partial_j\partial_t u_i.
	\label{eq.Gradient_strain_Kinetic_Energy}
\end{align}
where the $\rho$ is the mass density and $L_t$ is the characteristic inertial length. 

Solving the stationary condition problem $\delta S = 0$, with $S$ defined by eq. \eqref{eq.Minimization_Functional}, we can find the general equations of motion given by
Solving the stationary-action condition $\delta S=0$, with $S$
defined by eq.~\eqref{eq.Minimization_Functional}, gives the bulk
equations of motion
\begin{align}
	\rho
	\left(
	\delta_{ij}
	-
	L_t^2\delta_{ij}\partial_k\partial_k
	\right)
	\partial_t^2u_j
	&=
	\partial_j\sigma_{ij}
	-
	\partial_j\partial_k\tau_{ijk}
	+
	\frac{1}{2}
	\epsilon_{imn}
	\partial_m\partial_k
	m_{nk}.
	\label{eq.Gradient_Equation_Motion_General}
\end{align}
with $\tau,m$ defined in eqs. \eqref{eq.Gradient_Stress_Tensors}. In vector notation we can write
\begin{align}
	\rho
	\left(
	1
	-
	L_t^2\Delta
	\right)
	\partial_t^2 u
	&=
	\dvg \, \sigma
	-
	\dvg \, \dvg \, \tau
	+
	\frac{1}{2}
	\crl (\dvg \, m).
	\label{eq.Gradient_Equation_Motion_Vector}
\end{align}

\subsection{Dispersion analysis}

If we assume plane wave propagation of the form
\begin{align}
	u=  A_1 e^{\imag(k x + \omega t)},
	\label{eq.Plane_Waves1}
\end{align}
where $ A_1 $ is the (unknown) amplitude, $k$ the wavenumber and $\omega$ the angular frequency. In isotropic media, both longitudinal and transverse wave equations decouple. In both cases, the resulting 1D equation can be written in the common form as follows
\begin{align}
	\left[\partial_t ^2 - c_{x2}^2 \partial_x ^2  + c_{x4}^4 \partial_x ^4  - c_{xt}^2 \partial_t ^2\partial_x ^2\right]  u = 0 ,
	\label{eq.General_EQ_zeroI}
\end{align}
where $c_{x2}$ is the classical low-frequency phase velocity. The parameter $c_{xt}^2$ has dimensions of length squared and controls the mixed space--time derivative, the parameter $c_{x4}^4$ controls the fourth-order spatial derivative and together with $c_{xt}^2$ determines the high-frequency limit. 

We can obtain an expression for the phase velocity $(\omega/k)$ given by
\begin{align}
	v^2 =
	\frac{
		c_{x2}^2
		-
		c_{xt}^2\omega^2
		+
		\sqrt{
			\left(
			c_{x2}^2
			-
			c_{xt}^2\omega^2
			\right)^2
			+
			4c_{x4}^4\omega^2
		}
	}{2},
	\label{eq.General_Phase_vel_zeroI}
\end{align}
where we can identify the two limiting cases
\begin{align}
	\text{macroscale} \to	(v_{\rm macro})^2 = \lim_{\omega\to0}v^2 & = c_{x2}^2,&
	\text{microscale} \to 	(v_{\rm micro})^2 = \lim_{\omega\to\infty}v^2 &= \frac{c_{x4}^4}{c_{xt}^2}.&
	\label{eq.Phase_Velocity_limit}
\end{align}

Taking the ratio between the microscale and macroscale phase velocities we can write
\begin{align}
	(v_{\rm micro})^2
	=
	(v_{\rm macro})^2
\Lambda, \quad \text{with} \quad 	\Lambda
	=
	\frac{c_{x4}^4}
	{c_{x2}^2c_{xt}^2} .
	\label{eq.Phase_Velocity_Ratio}
\end{align}

\begin{figure}
	\begin{center}
		\includegraphics[width=1\textwidth]{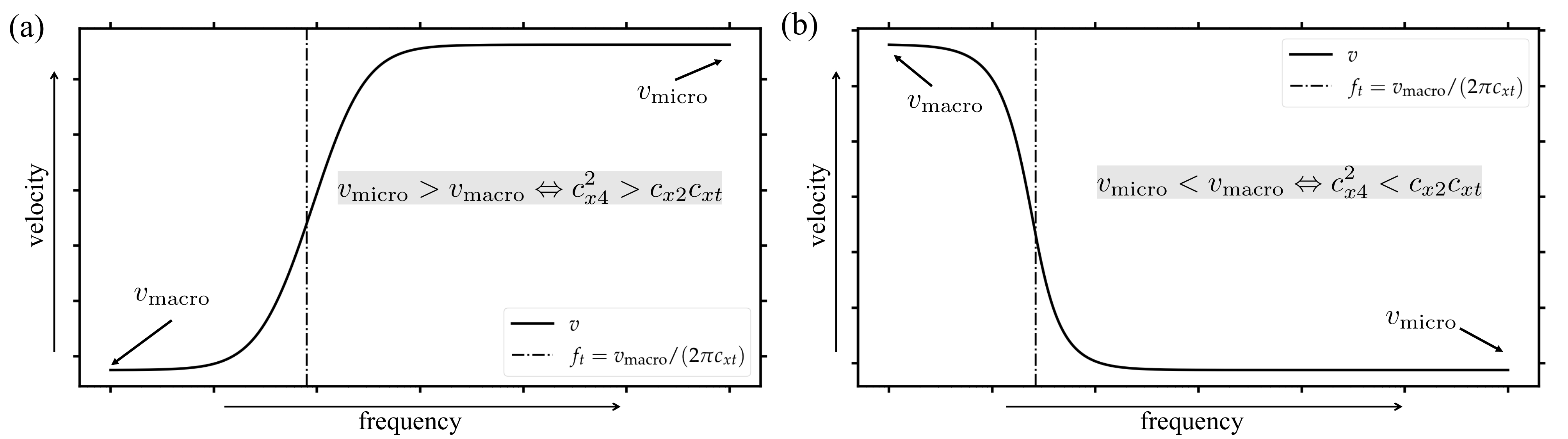}
		\caption{Shear velocity dispersion as a function of frequency for: (a) $ v_{\text{micro}} > v_{\text{macro}}$ and (b) $ v_{\text{micro}} < v_{\text{macro}}$.}
		\label{fig.micro_macro_excess}
	\end{center}
\end{figure}

Note that $\Lambda$ is a dimensionless parameter that controls the relative magnitude between the microscale and macroscale phase velocities. Figure~\ref{fig.micro_macro_excess} shows the two possible dispersion regimes, which are determined by the relative magnitudes of $c_{x4}$ and the product $c_{x2}c_{xt}$. Specifically,
\begin{align}
	c_{x4}^2
	>
	c_{x2}c_{xt}
	\quad
	\Longleftrightarrow
	\quad
	v_{\rm micro}
	>
	v_{\rm macro},
\end{align}
and
\begin{align}
	c_{x4}^2
	<
	c_{x2}c_{xt}
	\quad
	\Longleftrightarrow
	\quad
	v_{\rm micro}
	<
	v_{\rm macro}.
\end{align}

We can also write the non-dimensional (frequency-dependent) phase velocity as
\begin{align}
	\frac{v^2}
	{\left(v^{\rm macro}\right)^2}
	=
	\frac{
		1-\Omega^2
		+
		\sqrt{
			\left(1-\Omega^2\right)^2
			+
			4\Lambda\Omega^2
		}
	}{2}.
\end{align}
with
\begin{align}
	\Omega
	=
	\frac{c_{xt}\omega}{v^{\rm macro}},
\end{align}
and $\Lambda$ defined in eq. \eqref{eq.Phase_Velocity_Ratio}. The non-dimensional frequency $\Omega$ controls the transition between the macroscopic and microscopic regimes. For $\Omega\ll1$, wave propagation is dominated by the macroscopic phase velocity, whereas for $\Omega\gg1$ the influence of the higher-order constitutive terms becomes significant and the phase velocity gradually approaches its microscopic limit (see Fig.~\ref{fig.micro_macro_excess}).

For the higher-order gradient model introduced here,
\begin{align}
	\Omega
	=
	\frac{\omega L_t}{v_{\rm macro}}
	=
	\frac{2\pi f L_t}{v_{\rm macro}},
\end{align}
so that the characteristic transition frequency is given by
\begin{align}
	f_t
	\approx
	\frac{v_{\rm macro}}
	{2\pi L_t}.
	\label{eq.Transition_Frequency}
\end{align}

\subsubsection{1D shear waves}

In a 1D shear isotropic medium, the equation of motion reduces to
\begin{align}
	\rho
	\left(
	\partial_t^2u
	-
	L_t^2\partial_t^2\partial_x^2u
	\right)
	&=
	\mu\partial_x^2u
	-
	\left(
	\mu L_c^2 
	+
	\frac{\mu L_d^2}{4}
	\right)
	\partial_x^4u,
	\label{eq.Gradient_Wave_Equation_1D}
\end{align}
By comparison with the general wave equation (eq. \eqref{eq.General_EQ_zeroI}), we identify
\begin{align}
	c_{x2}^2
	&=
	\frac{\mu}{\rho},
	&
	c_{xt}^2
	&=
	L_t^2,
	&
	c_{x4}^4
	&=
	\frac{\mu}{\rho}
	\left(
 L_c^2 
	+
	\frac{L_d^2}{4}
	\right).
	\label{eq.Gradient_Shear_Coefficients}
\end{align}
The limiting phase velocities are
\begin{align}
	\left(v_S^{\rm macro}\right)^2 = \lim_{\omega\to0}v_S^2
	& =
	\frac{\mu}{\rho},
	& \left(v_S^{\rm micro}\right)^2 = \lim_{\omega\to\infty}v_S^2
	&=
	\frac{\mu}{\rho} \frac{1}{ L_t^2}
	\left(
 L_c^2 
	+
	\frac{L_d^2}{4}
	\right).
\end{align}

Using eq. \eqref{eq.Phase_Velocity_Ratio} we thus can write
\begin{align}
	\left(v_S^{\rm micro}\right)^2 = \left(v_S^{\rm macro}\right)^2 \Lambda_S \quad \text{with} \quad \Lambda_S =
	\frac{1}{L_t^2}
	\left(
	L_c^2+\frac{L_d^2}{4}
	\right).
	\label{eq.Micro_macro_shear_1D_waves}
\end{align}

\subsubsection{1D longitudinal waves}

In a 1D isotropic longitudinal medium, the equation of motion reduces to
\begin{align}
	\rho
	\left(
	\partial_t^2u
	-
	L_t^2\partial_t^2\partial_x^2u
	\right)
	&=
	\left(
	\lambda+2\mu
	\right)
	\partial_x^2u
	-
	\left(
	\lambda+2\mu
	\right)L_c^2
	\partial_x^4u .
	\label{eq:Gradient_P_Wave_Equation_1D}
\end{align}
By comparison with the general wave equation
\eqref{eq.General_EQ_zeroI}, we identify
\begin{align}
	c_{x2}^2
	&=
	\frac{\lambda+2\mu}{\rho},
	&
	c_{xt}^2
	&=
	L_t^2,
	&
	c_{x4}^4
	&=
	\frac{\lambda+2\mu}{\rho}
	L_c^2.
	\label{eq.Gradient_Longitudinal_Coefficients}
\end{align}
The limiting phase velocities are
\begin{align}
	\left(v_P^{\rm macro}\right)^2
	&=
	\lim_{\omega\to0}v_P^2 
	=
	\frac{\lambda+2\mu}{\rho},
&
	\left(v_P^{\rm micro}\right)^2
	&=
	\lim_{\omega\to\infty}v_P^2 
	=
	\frac{\lambda+2\mu}{\rho}
	\frac{L_c^2}{L_t^2}.
\end{align}
Using eq. \eqref{eq.Phase_Velocity_Ratio} we thus can write
\begin{align}
	\left(v_P^{\rm micro}\right)^2
	=
	\left(v_P^{\rm macro}\right)^2
	\Lambda_P \quad \text{with} \quad \Lambda_P =\frac{L_c^2}{L_t^2}.
		\label{eq.Micro_macro_longitudinal_1D_waves}
\end{align}

\subsection{Physical interpretation of higher-order gradient parameters}

The presented strain-gradient model contains three characteristic lengths: $L_s$, $L_c$, and $L_t$. Each parameter controls a different higher-order deformation mechanisms and influences wave propagation in a different way.

\subsubsection{The inertial characteristic length $L_t$}

The characteristic length $L_t$ appears in the kinetic-energy density (eq.~\eqref{eq.Gradient_strain_Kinetic_Energy}) and therefore characterizes gradient inertia. It controls the dynamic resistance of the medium to spatially varying accelerations and determines the transition between the low-frequency (macroscopic) and high-frequency (microscopic) wave-propagation regimes, through the characteristic frequency $f_t$ (see eq. \eqref{eq.Transition_Frequency}) at which higher-order inertial effects become significant.

\subsubsection{The elastic characteristic length $L_c$}

The characteristic length $L_c$ enters the elastic-energy density through the strain-gradient contribution. It controls the energetic resistance to spatial variations of strain and therefore contributes to the fourth-order elastic response. Since both longitudinal and transverse waves involve nonzero strain gradients, $L_c$ affects the dispersion of both P and S waves.

\subsubsection{The curvature characteristic length $L_d$}

The characteristic length $L_d$ enters the elastic-energy density through the curvature contribution. It controls the energetic resistance to spatial variations of local rotations. Since purely longitudinal motion does not involve local rotation, $L_d$ contributes only to transverse-wave dispersion. Using eq. \eqref{eq.Micro_macro_shear_1D_waves} and eq. \eqref{eq.Micro_macro_longitudinal_1D_waves} we can write
\begin{align}
		\Lambda_P
	&=
	\frac{L_c^2}{L_t^2},
	&
	\Lambda_S
	&=
	\frac{1}{L_t^2}
	\left(
	L_c^2+\frac{L_d^2}{4}
	\right)
	=
	\Lambda_P+\frac{L_d^2}{4L_t^2}.
\end{align}
As a consequence
\begin{align}
	\Lambda_S\geq\Lambda_P,
\end{align}
and the model predicts that the normalized microscopic-to-macroscopic velocity ratio for S waves cannot be smaller than that for P waves. This prediction will be tested against laboratory measurements in the following sections.

\subsubsection{The takeaway}

The proposed strain-gradient model does not attempt to explicitly describe the underlying microstructure of the material under study. Instead, it captures the mechanical consequences that the microstructure has on the macroscopic response through higher-order deformation measures. In this sense, the model should be understood as an effective continuum description of wave propagation rather than as a direct representation of the physical morphology.

The characteristic lengths $L_c$, $L_d$, and $L_t$ should not be interpreted as direct geometric measures of grains, pores, fractures, or inclusions. They simply quantify the effective mechanical influence that any microstructural properties of the material have on the observable dynamic response. The characteristic length $L_c$ measures the additional stiffness associated with strain gradients, $L_d$ measures the additional stiffness associated with gradients of local rotations, and $L_t$ characterizes the inertial length governing the transition between macroscopic and microscopic wave propagation. Together, these parameters determine the dispersive behavior of the medium.

\begin{figure}
	\begin{center}
		\includegraphics[width=1\textwidth]{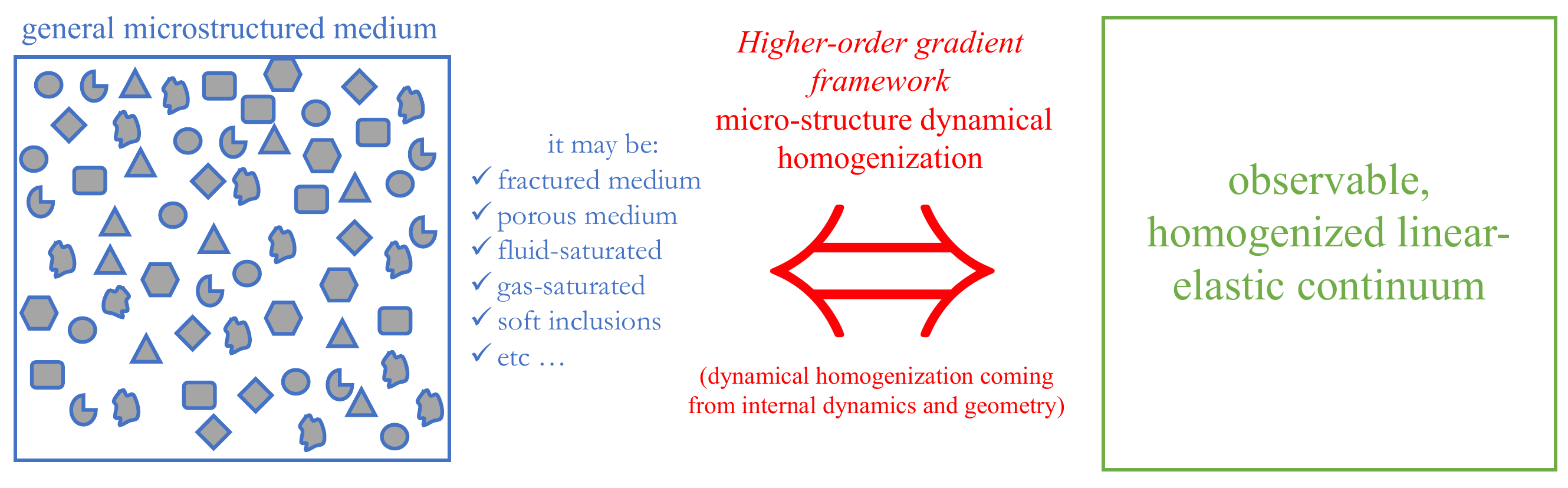}
		\caption{Different microstructural configurations (e.g., cracks, pores with fluid/gas filled regions, or hard/soft inclusions) can be mapped into effective dynamical macroscopic responses within the introduced framework.}
		\label{Fig.relaxed_micro_homogenization}
	\end{center}
\end{figure}

A direct consequence is that different microstructural configurations, such as fractured media, porous materials containing fluids or gases, or materials with soft or stiff inclusions, may exhibit indistinguishable dispersion curves (see Fig.~\ref{Fig.relaxed_micro_homogenization}). As a consequence, the proposed framework characterizes the effective mechanical properties of the microstructure rather than any specific physical property/geometry.

\section{Experimental validations}
\label{se_GeneralExperimentalValidations}

\subsection{Frequency-dependent seismic velocities in fully saturated tight sandstones}

Tight sandstone reservoirs are compacted sedimentary formations that are capable of storing hydrocarbons. They are increasingly being recognized as unconventional energy resources \citep{Zou2010,ZOU2015,JIA2012139,zoback2019unconventional,li2015review}. In contrast to conventional reservoirs, tight sandstones reservoirs have experienced prolonged mechanical compaction and cementation, which reduce pore space and increase rock stiffness as mineral precipitates progressively fill intergranular voids \citep{morad2010impact,wilson2014influence}, which makes fluid flow  much more complicated and reduces the hydrocarbon storage capacity. As a consequence, these unconventional characteristics introduce new challenges for seismic exploration and reservoir imaging, since tight formations require more sophisticated approaches due to their complex pore geometry and weak seismic responses \citep{zou2012tight,Mavko2009,Avseth2010}. 

Laboratory-based rock physics experiments are essential in order to link reservoir rock properties to their corresponding seismological observations. Without this, the quantitative  interpretation of geophysical data and reservoir characterization becomes even harder \citep{Avseth2010,Mavko2009}. However, even when the two observations are available, the direct link between laboratory experiments and seismological observations is highly non-linear. For instance, fluid-saturated rocks very often exhibit a strong frequency-dependent behavior that results in dispersion and attenuation effects across different high frequency bands \citep{muller2010,chapman2003frequency}. These frequency bands are, in general, much higher in omparison to the frequency bands that seismological experiments are performed. Therefore, the discrepancy between ultrasonic and seismic frequencies must be carefully considered when interpreting experimental data. 

In general, seismic wave propagation in such complex reservoir media is described using Biot's theory of poroelasticity \cite{Biot1956,Biot1956b}, and/or its extensions/modifications such as the Biot-Stoll model \cite{chotiros2004broadband,Attenborough1986,holland1988biot,badiey1998geology,shin2013non}. However, despite its predictive power, the Biot–Stoll model possesses a total of thirteen parameters (grain density, grain bulk modulus, fluid density, fluid bulk modulus, porosity, viscosity, permeability, pore size, tortuosity, static frame shear modulus, grain Poisson's ratio, asymptotic fluid contribution and bulk relaxation frequency), and in some cases, it is not even capable of reproducing detailed observations of shear wave propagation in fluid saturated sands \cite{brunson1984shear}. 

In order to reproduce shear velocity dispersion observations, further modifications to the Biot's model have been proposed. For instance, the Biot modified gap stiffness model (BIMGS) \cite{kimura2006frame,kimura2008experimental,kimura2013shear} possesses, in addition to the thirteen parameters of the Biot's model, a total of four more parameters (\textit{Hertz–Mindlin shear modulus, the maximum gap stiffness term of the frame shear modulus, the aspect ratio and the relaxation frequency}), for a total of seventeen parameters. The large amount of parameters required to model seismic wave propagation using these theories, as one can expect, makes them unpractical for seismological purposes. In contrast, as we will next see, the relaxed-micromorphic model is able to fully reproduce the experimental data with a reduced number of three parameters only.  

\subsubsection{Experimental verification}

In a recent study by \cite{yin2017pressure}, a tight sandstone sample characterized by a crack--pore microstructure has been investigated under different saturation conditions, including dry (nitrogen), brine, and glycerin. The sample was obtained from a drill core at depths of 4458--4466 m within the Xujiahe Formation in the northeastern Sichuan Basin, China \citep{Xu2008}. Cylindrical specimens with a diameter of 38 mm and a length of 70 mm were prepared for laboratory measurements. The mineral composition is dominated by quartz, feldspar, and calcite, with additional contributions from clays and other interstitial phases. The effective elastic moduli of the mineral matrix were estimated using the Voigt--Reuss--Hill averaging scheme \citep{hill1952elastic}. 

The investigated sandstone exhibits very low permeability ($6.3 \times 10^{-17}$~m$^2$) and a porosity of approximately 8.9$\%$, measured using helium as the reference fluid. In such tight formations, pore structure plays a more critical role than porosity alone in controlling elastic properties \citep{smith2009rock}. To better characterize the pore system, a 3~mm diameter sub-sample was imaged using high-resolution 3D X-ray microscopy, enabling detailed analysis of pore geometry and connectivity. 

Experimental observations reveal pronounced dispersion and attenuation associated with the transition between undrained and unrelaxed regimes in fluid-saturated conditions (brine and glycerin). These effects are progressively reduced with increasing effective pressure. A squirt-flow model incorporating a dual-porosity framework was used to interpret the measurements. Although minor discrepancies are observed, the model successfully reproduces the overall trends of P wave velocity dispersion and particularly at low effective pressures \citep{mavko1980velocity,Mavko2009}. 

The results obtained by \cite{yin2017pressure} suggest that, in tight sandstones with crack--pore systems, dispersion and attenuation are primarily governed by squirt-flow mechanisms, where fluid exchange between cracks and pores leads to frequency-dependent elastic stiffening \citep{chapman2003frequency}. To investigate the effect of cracks on the wave propagation response, \cite{ba2023effect} measured porosity, permeability, and ultrasonic velocities at different differential pressures, and performed forced-oscillation measurements on core samples and a theoretical model based on poroelastic theory with a single set of penny-shaped cracks \cite{zhang2019modeling} was proposed to interpret results. 

\begin{figure}
	\begin{center}
		\includegraphics[width=1\textwidth]{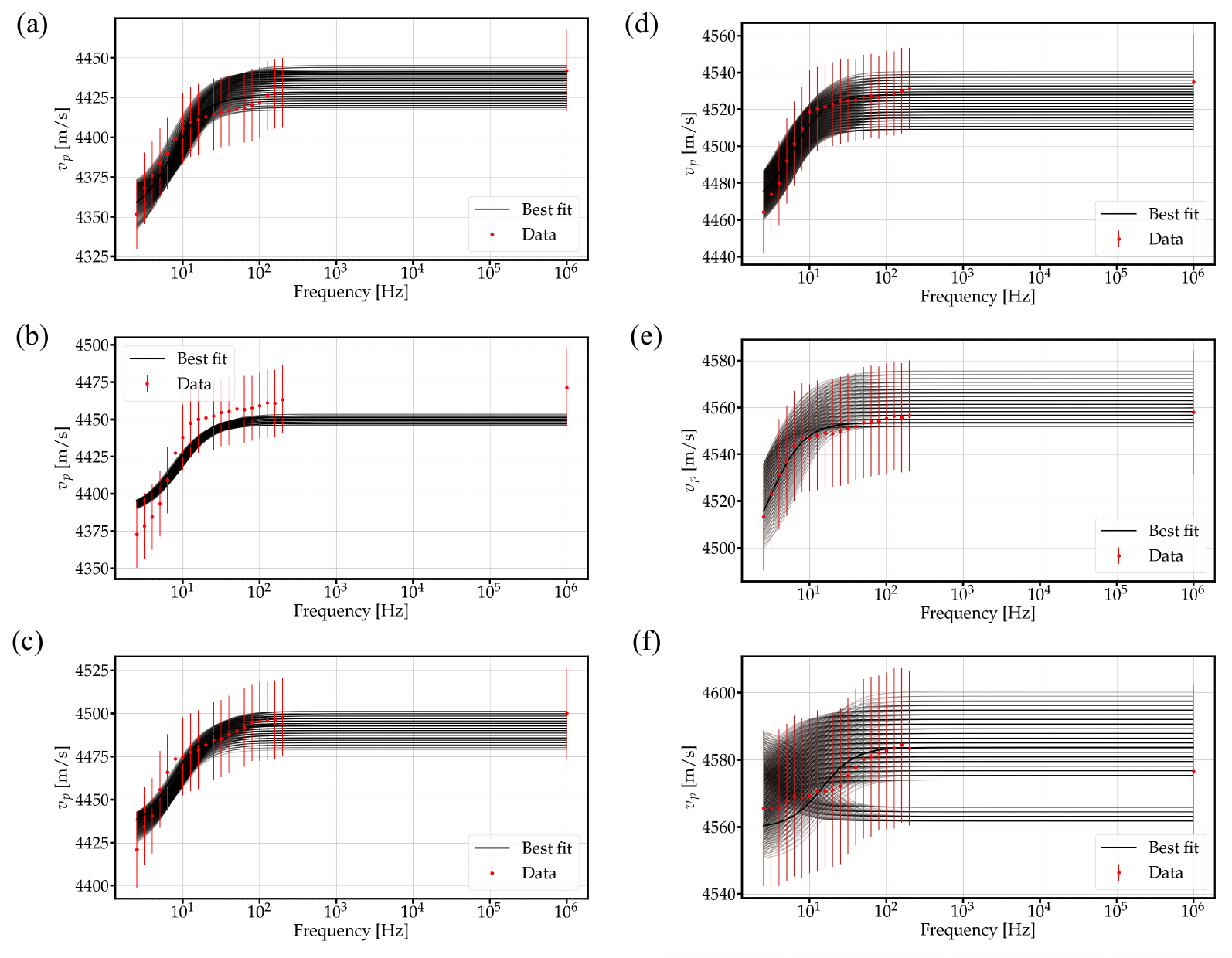}
		\caption{Frequency-dependent longitudinal-wave velocity data (from \cite{yin2017pressure}) with error bars and corresponding models obtained from all admissible parameter combinations that satisfy the observations.}
		\label{Fig.experiments_Vp}
	\end{center}
\end{figure}

We can observe the level of mathematical complexity needed to properly describe such difficult experimental scenarios requires a high level of sophistication that it is not available from seismological scenarios/experiments. In order to reduce these limitations, we next compare the higher-order gradient model predictions to experimental data from \cite{yin2017pressure}. To do so, we perform a grid search inversion for the required elastic parameters. Figs. \ref{Fig.experiments_Vp} shows the obtained results for P waves.
The higher-order gradient model accurately reproduces the experimentally observed frequency-dependent P-wave velocities for all considered confining pressures (Fig. \ref{Fig.experiments_Vp}). At lower confining pressures, the separation between the macroscopic and microscopic phase velocities is more pronounced, indicating stronger dispersion than at higher pressures. The characteristic lengths $L_t,L_c$, retain the same ratio as  $v_P^{\rm macro},v_P^{\rm micro}$. In this sense, the individual values of $L_t$ and $L_c$ should not be interpreted independently, but in conjunction with the transition frequency $f_t$ (see eq.  \eqref{eq.Transition_Frequency}), which determines at which frequency the transition from the macroscopic to the microscopic regime occurs (see Fig. \ref{fig.micro_macro_excess}). In this sense, these characteristic lengths are related to the transition regime and not to materials parameters. We can observe that he admissible models become progressively more tightly clustered with increasing confining pressure. 

Obtained numerical values are reported in Table \ref{tab:Pwave_parameters}. We can observe that the transition frequency generally decreases with increasing confining pressure, shifting the transition between the macroscopic and microscopic regimes towards lower frequencies. As a consequence, the phase-velocity curves become flatter over the measured frequency range. At higher confining pressures, the transition occurs outside the experimental frequency band, making the inversion unable to constrain its location. In this regime, the observed response becomes practically indistinguishable from that predicted by classical linear elasticity.
\begin{table}
	\centering
	\caption{Best-fitting parameters obtained from the higher-gradient model for longitudinal (P) waves. The transition frequency $f_t$ is estimated using eq.~\eqref{eq.Transition_Frequency}.}
	\label{tab:Pwave_parameters}
	\begin{tabular}{ccccccc}
		\toprule
		Pressure [MPa] &
		$v_{\rm micro}$ [m/s] &
		$v_{\rm macro}$ [m/s] &
		$L_t$ [m] &
		$L_c$ [m] &
		$f_t$ [Hz] &
		$\chi^2$ \\
		\midrule
		5  & 4425.70 & 4350.69 & 98.31  & 100.00 & 7.04  & 1.54 \\
		7  & 4451.38 & 4390.82 & 81.87  & 83.00  & 8.54  & 6.14 \\
		10 & 4492.85 & 4431.72 & 98.64  & 100.00 & 7.15  & 1.46 \\
		15 & 4527.98 & 4466.38 & 119.41 & 121.05 & 5.95  & 0.67 \\
		20 & 4553.50 & 4491.55 & 228.43 & 231.58 & 3.13  & 0.17 \\
		25 & 4583.65 & 4559.66 & 49.74  & 50.00  & 14.59 & 0.50 \\
		\bottomrule
	\end{tabular}
\end{table}

\begin{figure}
	\begin{center}
		\includegraphics[width=1\textwidth]{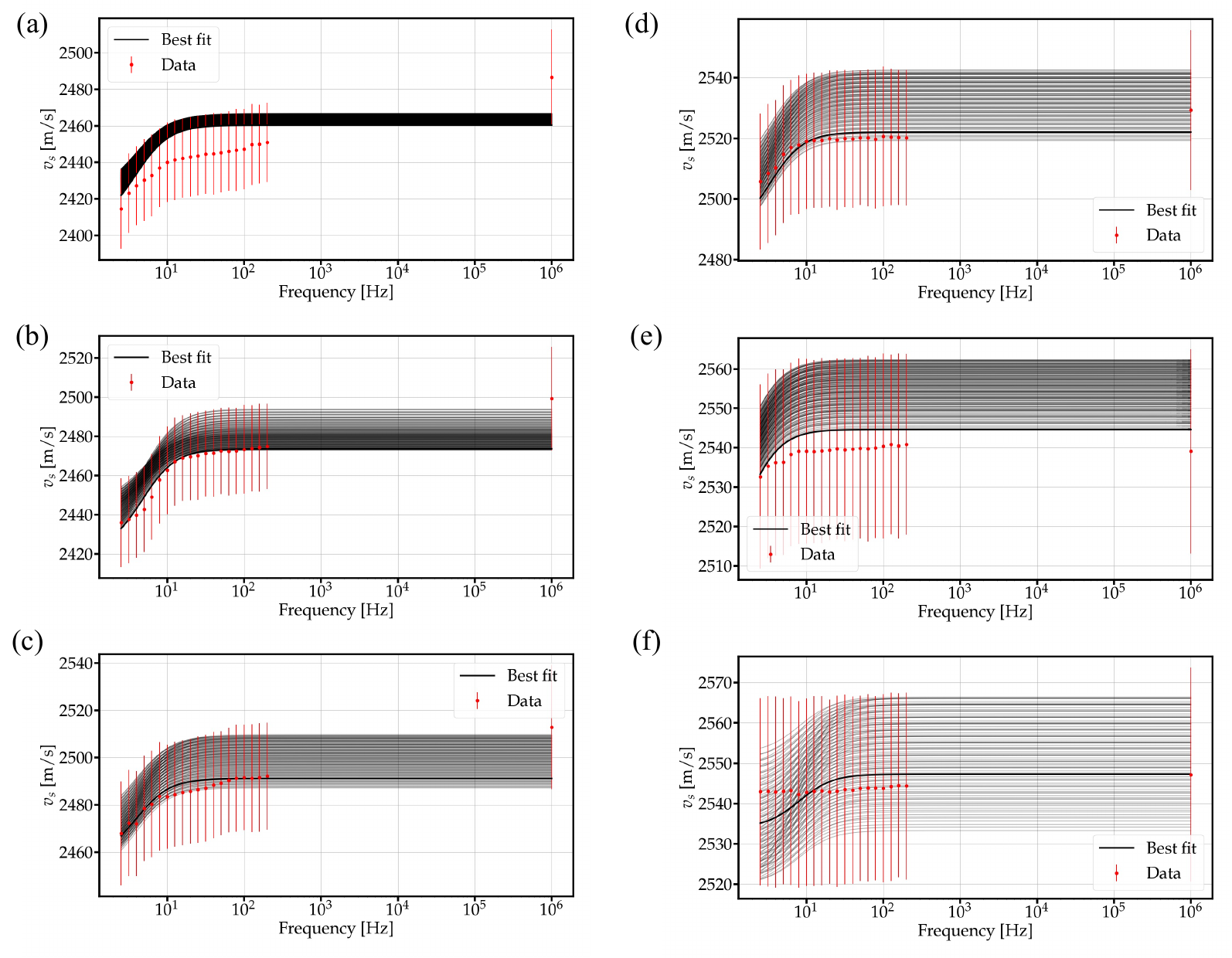}
		\caption{Frequency-dependent shear-wave velocity data (from \cite{yin2017pressure}) with error bars and corresponding models obtained from all admissible parameter combinations satisfying the experimental constraints.}
		\label{Fig.experiments_Vs}
	\end{center}
\end{figure}

Fig.~\ref{Fig.experiments_Vs} shows the corresponding inversion results for shear waves. During the inversion, the characteristic lengths $L_t$ and $L_c$ are fixed to the values independently obtained from the corresponding P-wave inversion. Thus the only additional inversion parameter is the characteristic length $L_d$, associated with the curvature contribution.

Obtained numerical values are reported in Table~\ref{tab:Swave_parameters}. We can observe that the characteristic length $L_d$ decreases with increasing confining pressure and eventually approaches zero. This indicates that the additional curvature contribution becomes progressively less important as the material response approaches the behavior of classical linear elasticity. In other words, the strain-gradient contribution alone becomes sufficient to reproduce the observed shear-wave dispersion at high confining pressures.

Furthermore, the transition frequency is consistently lower for S waves compared to P waves. Consequently, the transition between the macroscopic and microscopic regimes occurs over a lower frequency range for shear waves.
\begin{table}
	\centering
	\caption{Best-fitting parameters obtained from the higher-gradient model for transverse (S) waves. The transition frequency $f_t$ is estimated using eq.~\eqref{eq.Transition_Frequency}, where the characteristic lengths $L_t,L_c$ are taken from the corresponding P-wave inversion.}
	\label{tab:Swave_parameters}
	\begin{tabular}{ccccccc}
		\toprule
		Pressure [MPa] &
		$v_{\rm micro}$ [m/s] &
		$v_{\rm macro}$ [m/s] &
		$L_d$ [m] &
		$f_t$ [Hz] &
		$\chi^2$ \\
		\midrule
		5 & 2460.34 & 2405.00 & 21.32 & 3.90 & 7.34 \\
		7 & 2473.67 & 2421.28 & 20.69 & 4.71 & 1.16 \\
		10 & 2490.77 & 2456.03 & 5.26  & 3.97 & 1.03 \\
		15 & 2521.84 & 2487.11 & 4.47  & 3.32 & 1.03 \\
		20 & 2544.62 & 2510.00 & 0.00  & 1.75 & 0.70 \\
		25 & 2547.30 & 2533.97 & 0.00  & 8.11 & 0.57 \\
		\bottomrule
	\end{tabular}
\end{table}

\section{Discussion}
\label{sec.Discussion}

\subsection{Frequency-dependent seismic responses}

The inversion reveals that both the inertial characteristic length $L_t$ and the strain-gradient characteristic length $L_c$ are common to longitudinal and transverse waves. This suggests that gradient inertia and strain-gradient elasticity represent fundamental dynamic mechanisms acting on both volumetric and shear deformation. In contrast, only the curvature characteristic length $L_d$ is specific to shear waves, indicating that rotational deformation provides an additional constitutive mechanism affecting exclusively transverse-wave propagation.

The inversions also show that the contribution of the curvature term decreases with increasing confining pressure, eventually becoming negligible within the experimental uncertainty. Physically, this indicates that the rotational degrees of freedom become progressively less relevant as grain contacts stiffen under compression, and the observed response approaches that of classical elasticity.

Another interesting observation is that the transition frequency is consistently lower for shear waves than for compressional waves. Since the characteristic inertial length $L_t$ is common to both wave types, this difference follows directly from the smaller macroscopic shear-wave velocity. Consequently, shear waves probe the same internal length scale over a lower frequency range and are therefore expected to be more sensitive to microstructural effects within laboratory experiments.

Finally, the inversion confirms that the higher-order gradient framework reproduces the observed laboratory dispersion using only a small number of effective constitutive parameters. Rather than describing the detailed morphology of pores or fractures, these parameters characterize the effective dynamical response of the medium. In this sense, the proposed model provides a dynamic homogenization framework capable of representing complex poroelastic behavior without explicitly modeling the underlying pore-fluid interactions.

This behavior differs fundamentally from Biot's poroelastic theory, where wave dispersion originates from the relative motion between the pore fluid and the solid skeleton and is governed by fluid viscosity and permeability \citep{Carcione2015}. In contrast, the proposed higher-order gradient framework reproduces the observed dispersion using effective elastic and inertial parameters, without introducing a pore fluid or dissipative mechanisms. Consequently, the observed laboratory dispersion does not uniquely imply Biot-type fluid description. Instead, it demonstrates that the same macroscopic wavefield can be represented by an effective higher-order constitutive description based on characteristic length scales associated with gradient elasticity and gradient inertia.

\subsection{P--wave dispersion: comparison with classical Biot's model}

For 1D longitudinal-wave propagation, the solid displacement $u(x,t)$ and the relative fluid displacement $w(x,t)$ are both parallel to the direction of propagation. Using the same notation and sign convention adopted for the shear-wave equations, Biot's longitudinal equations can be written as \citep{bourbie1987acoustics,van2013multi,Carcione2015}
\begin{align}
	\begin{aligned}
		\rho \,\partial_t^2 u
		-
		\rho_f\,\partial_t^2 w
		&=
		H\,\partial_x^2 u
		-
		\alpha M\,\partial_x^2 w,
		\\
		\rho_f\,\partial_t^2 u
		-
		m\,\partial_t^2 w
		-
		\frac{\eta F}{\kappa}\,\partial_t w
		&=
		\alpha M\,\partial_x^2 u
		-
		M\,\partial_x^2 w,
	\end{aligned}
	\label{eq.Biot_P_Wave_Compact}
\end{align}
with
\begin{align}
	H
	=
	K_b+\frac{4}{3}\mu+\alpha^2M
	=
	\lambda+2\mu+\alpha^2M,
\end{align}
where $u$ is the longitudinal displacement of the solid skeleton and $w$ is the is the relative displacement of the pore fluid with respect to the solid. The parameter $\phi$ is the porosity, $\rho_s$ is the grain density, $\rho_f$ is the pore-fluid density, $\alpha_\infty$ is the tortuosity, $\eta$ is the dynamic viscosity of the pore fluid, $\kappa$ is the permeability, and $F$ is the viscous correction factor.

In the inviscid limit, the viscous drag term vanishes, i.e.,
\begin{align}
	\frac{\eta F}{\kappa}\partial_t w=0,
\end{align}
and eq. \eqref{eq.Biot_P_Wave_Compact}, can be written in a compact form as follows
\begin{align}
	\left[
	\partial_t^4
	-
	c_{xt}^2\,\partial_t^2\partial_x^2
	+
	c_{x4}^4\,\partial_x^4
	\right](u,w) 
	=
	0,
	\label{eq.Biot_P_Inviscid_General}
\end{align}
where
\begin{align}
	c_{xt}^2
	&=
	\frac{
		\rho M
		+
		Hm
		-
		2\alpha M\rho_f
	}{
		\rho m-\rho_f^2
	},
	&
	c_{x4}^4
	&=
	\frac{
		M\left(H-\alpha^2M\right)
	}{
		\rho m-\rho_f^2
	}.
\end{align}

Note that we can factor eq. \eqref{eq.Biot_P_Inviscid_General} as follows
\begin{align}
	\left[
	\partial_t^2
	-
	c_{P+}^2\partial_x^2
	\right]
	\left[
	\partial_t^2
	-
	c_{P-}^2\partial_x^2
	\right]u
	=
	0,
\end{align}
with
\begin{align}
	c_{P\pm}^2
	=
	\frac{
		c_{xt}^2
		\pm
		\sqrt{
			c_{xt}^4
			-
			4c_{x4}^4
		}
	}{2}.
	\label{eq.Biot_Decoupled_Velocities}
\end{align}
Therefore we can write
\begin{align}
	c_{xt}^2
	&=
	c_{P+}^2+c_{P-}^2,
	&
	c_{x4}^4
	&=
	c_{P+}^2c_{P-}^2.
\end{align}
where $v_{P+}$ and $v_{P-}$ denote the fast and slow compressional-wave velocities, respectively. 

This factorization shows that the fourth-order inviscid Biot equation is the product of two classical second-order wave operators. The coupled solid--fluid system therefore supports two distinct compressional eigenmodes, namely the fast and slow P waves, with constant velocities $c_{P+}$ and $c_{P-}$, respectively. Thus, under the inviscid, homogeneous, and constant-parameter assumptions, each compressional branch is nondispersive.

This differs fundamentally from the strain-gradient model introduced in this work. In the strain-gradient formulation, a single compressional-wave branch changes continuously between its macroscopic and microscopic velocity limits. In inviscid Biot theory, by contrast, $c_{P+}$ and $c_{P-}$ correspond to two distinct wave modes rather than to the low- and high-frequency limits of one branch.

This factorization immediately shows that the characteristic parameters of the inviscid Biot model cannot be directly identified with the characteristic lengths of the proposed strain-gradient model. In the inviscid Biot formulation, the fourth-order operator simply combines two distinct compressional eigenmodes (fast and slow P waves), each governed by a classical second-order wave equation. In contrast, the strain-gradient model contains a single compressional branch whose velocity evolves continuously between its macroscopic and microscopic limits as the frequency increases.

Consequently, the physical origin of dispersion in both theories is fundamentally different. In classical Biot theory, dispersion arises from the relative motion between the fluid and solid phases once viscous or frequency-dependent hydrodynamic effects are retained. By contrast, the proposed strain-gradient model predicts dispersion through higher-order elastic and inertial effects, without requiring an explicit fluid phase. Therefore, the model provides a more general continuum description in which wave dispersion results from the effective mechanical consequences of the microstructure rather than from a specific fluid--solid interaction mechanism.

\subsection{S--wave dispersion: limitations of classical Biot's models}

It is well known that Biot's theory encounters important challenges for modeling S-wave dispersion \citep[e.g.][]{ba2023effect}. This limitation is related to fluid–solid coupling assumed in Biot's theory, for which S-waves are weakly sensitivity.  The 1D Biot's shear equations of motion can be written as follows \citep{bourbie1987acoustics,van2013multi,Carcione2015}
\begin{align}
	\begin{aligned}
		\left[(1-\phi)\rho_s+\phi\rho_f\right]\partial_t^2 u-\rho_f\partial_t^2 w & =\mu\partial_x^2 u,
		\\
		\rho_f\partial_t^2 u-\frac{\alpha_\infty\rho_f}{\phi}\partial_t^2 w-\frac{\eta F}{\kappa}\partial_t w& =0,
	\end{aligned}
	\label{eq.Biot_Equation_Explicit}
\end{align}
where $u$ is the transverse displacement of the solid skeleton and $w$ is the relative displacement of the pore fluid with respect to the solid. The parameter $\phi$ is the porosity, $\rho_s$ is the grain density, $\rho_f$ is the pore-fluid density, $\alpha_\infty$ is the tortuosity, $\eta$ is the dynamic viscosity of the pore fluid, $\kappa$ is the permeability, and $F$ is the viscous correction factor. 

In the inviscid limit, the viscous drag term vanishes, i.e.,
\begin{align}
	\frac{\eta F}{\kappa}\partial_t w=0,
\end{align}
and eq. \eqref{eq.Biot_Equation_Explicit} can be written as follows
\begin{align}
	\left[ \partial_t ^2  - c_{x2}^2 \partial_x ^2 \right]  (u,w) = 0 ,
	\label{eq.General_EQ}
\end{align}
where the material parameter is given by
\begin{align}
	c_{x2}^2=\frac{\mu\alpha_\infty}{\alpha_\infty\left[(1-\phi)\rho_s+\phi\rho_f\right]-\phi\rho_f}.
\end{align}
Therefore, in the inviscid limit, Biot's theory predicts a single shear-wave velocity that is independent of frequency. In other words, the model does not contain a mechanism capable of producing intrinsic S-wave dispersion. This explains why the inviscid Biot formulation cannot reproduce the frequency-dependent shear-wave velocities commonly observed in laboratory experiments.

\subsection{Implications for natural gas reservoir imaging}

\begin{figure}
	\begin{center}
		\includegraphics[width=1\textwidth]{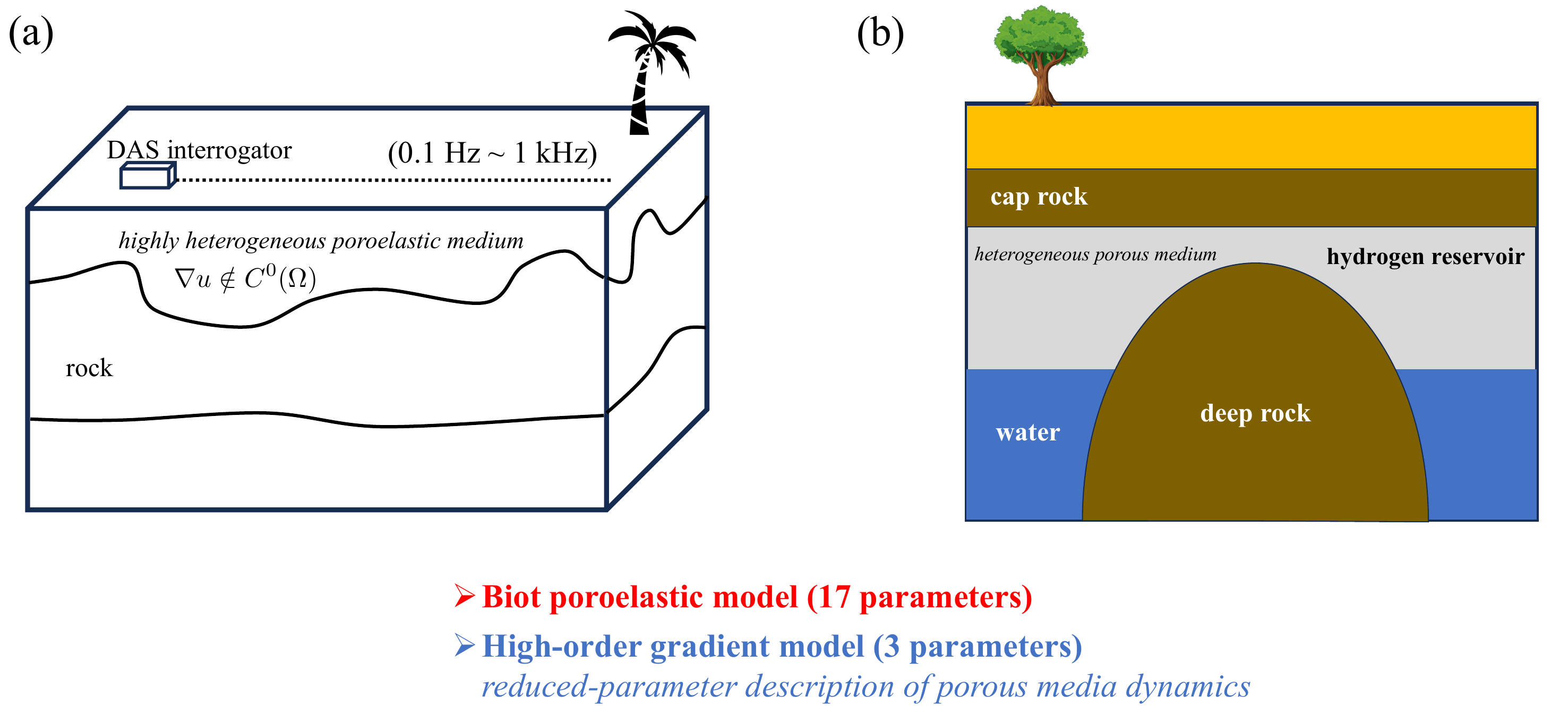}
		\caption{(a) Dynamic homogenization of a poroelastic medium to model seismic signals measured with DAS. (b) Seismic imaging of a geologic hydrogen reservoir.}
		\label{fig.deep_earth}
	\end{center}
\end{figure}

White (or geologic) hydrogen is molecular hydrogen naturally generated in the subsurface through geological processes such as serpentinization of ultramafic rocks, radiolysis of water, iron oxidation, hydrothermal reactions, and deep mantle degassing \citep{gaucher2020new,scott2021exploring}. It is encountered in faults, sedimentary basins, fracture zones, and porous geological formations \citep{jackson2024natural}. Unlike fossil fuels—which formed over millions of years, white hydrogen is continuously produced in many geological environments and emits no carbon when burned, making it a highly attractive potentially renewable, carbon-free energy source (see Fig. \ref{fig.deep_earth}--b).

As a naturally occurring form of hydrogen, it may play a key role in the future hydrogen economy, supporting applications ranging from fuel cells to industrial processes and power generation. Its potential to complement other hydrogen sources (green, blue, and turquoise hydrogen) makes it an attractive candidate for large-scale decarbonization. However, despite recent scientific interest, the exploration of white hydrogen remains in its infancy. Considerable geophysical, geological, and technological advances are required to assess its global abundance, extractability, and long-term sustainability \citep{rigollet2022natural}.

\subsubsection{Seismic imaging}

Subsurface resources such as hydrocarbons or groundwater are commonly explored using \textit{seismic tomography}, a method that reconstructs images of the Earth's interior by solving the inverse problem using seismic wavefields. Conventional tomographic studies for oil and gas, fundamentally rely on linear elastic wave theory. However, hydrogen-bearing formations are typically partially saturated with hydrogen and water. In situ, hydrogen occurs as highly compressible bubbles that strongly influence wave propagation. The presence of gas bubbles leads to substantial attenuation and dispersion, effects that become particularly pronounced in porous or fractured media where poroelastic interactions are dominant. Seismic imaging of white hydrogen reservoirs requires accounting for poroelastic wave propagation, where the interaction between the solid matrix and pore fluids leads to complex seismic signatures.

\subsubsection{Seismic imaging with higher-order gradient elasticity}

The preceding analysis suggests that higher-order gradient elasticity provides an effective continuum description capable of reproducing the principal dispersive behavior observed in porous media using only three characteristic length parameters. In contrast, classical poroelastic formulations, such as the Biot--Stoll model, require a considerably larger number of constitutive parameters, many of which are difficult to estimate independently from seismic observations \citep[e.g.][]{kimura2006frame,kimura2008experimental,kimura2013shear}.

The proposed framework captures the effective mechanical behavior on wave propagation through higher-order elastic and inertial contributions. The model thus provides a reduced-order representation of poroelastic dispersion that may substantially decrease the dimensionality of the inverse problem while preserving the dominant dispersive features relevant for seismic wave propagation.

This reduction in the number of unknown parameters has the potential to improve the robustness and stability of seismic inversion procedures, particularly in applications where the physical properties of the pore fluid and the microstructure are poorly constrained. One important example is the seismic exploration of naturally occurring hydrogen reservoirs, where the distribution of hydrogen, water, fractures, and pore geometry is largely unknown. Within this context, the proposed higher-order gradient framework offers a promising alternative to classical poroelastic descriptions by replacing numerous poorly constrained microscopic parameters with a small number of effective constitutive quantities that can be estimated directly from observed wave dispersion.

\section{Conclusions}
\label{sec.Conclusions}

We have presented a higher-order gradient constitutive framework for modeling frequency-dependent seismic wave propagation in porous media. The proposed formulation extends classical strain-gradient elasticity by incorporating an additional curvature contribution associated with rotational deformation, allowing independent descriptions of longitudinal and transverse wave dispersion within a unified continuum model.

A general dispersion relation was derived together with its corresponding macroscopic and microscopic velocity limits. These limits naturally define characteristic inertial and elastic length scales that govern the transition between classical and higher-order wave propagation. We showed that the characteristic lengths should be interpreted as effective dynamical parameters describing the mechanical influence of unresolved microstructure rather than as direct geometric measures of pores, fractures, or inclusions.

Comparison with classical Biot poroelasticity demonstrated that the proposed framework reproduces frequency-dependent wave velocities using only a small number of effective constitutive parameters. In particular, the model successfully reproduces laboratory measurements of both P- and S-wave dispersion while avoiding the large number of poorly constrained parameters required by conventional poroelastic formulations. The inversion further shows that the inertial characteristic length $L_t$ and the strain-gradient characteristic length $L_c$ are common to both wave types, whereas the curvature characteristic length $L_d$ provides an additional constitutive mechanism acting only on shear-wave propagation.

Finally, the proposed interpretation suggests that higher-order gradient elasticity may be understood as an effective dynamic homogenization framework for poroelastic wave propagation. This provides a physically interpretable framework for seismic wave propagation and offers a promising alternative for large-scale inversion and imaging problems in Earth sciences, including applications to unconventional reservoirs and natural hydrogen exploration.

\section{Acknowledgments}

R.A. acknowledges constructive conversations with Fabian Bonilla. 

\section{Data availability}

Data used in Section \ref{se_GeneralExperimentalValidations} can be downloaded from the supporting information of \cite{yin2017pressure}. No new experimental data were generated in this study. All equations required to reproduce the theoretical results are provided in the manuscript.

\section{Conflict of interest}

The authors have no conflicts of interest.

\footnotesize

\bibliographystyle{apalike}
\bibliography{Biblio}

\end{document}